# A Mechatronics view at nerve conduction

Is that the way we feel, memorize, think and act?

Jörg P. Kotthaus,

Faculty of Physics and Center for Nanoscience (CeNS),

LMU Munich, Geschwister-Scholl-Platz 1, D 80539 Munich, Germany

## Abstract

Stimulated by ongoing discussions about the relevance of mechanical motion in the propagation of nerve signals capillary waves of water-based electrolytes in elastic tubular systems are considered as an essential ingredient. Their propagation velocities, controlled by the elastic properties and geometry of the neuron membrane as well as the density of the confined electrolyte, are shown to very well match observed nerve conduction velocities. As they experience little damping and exhibit non-linear behavior they can behave soliton-like. The orientation of water dipoles by the high electric fields up to about $10^7$ V/m in the nanometer scale thick Debye layer adjacent to the elastic neuron wall causes radial forces that modulate the diameter of the nerve cell with a change of the voltage bias across the cell membrane, caused, e. g., by local injection of ions. Acting like an electrically driven peristaltic pump the soma thus can launch capillary waves into the axon which also transport neutral dipolar current pulses modulated the varying voltage bias. At the synaptic end of the axon these dipolar current pulses can cause voltage changes and initiate chemical signaling. In contrast to the traditional Hodgkin-Huxley model of nerve conduction the proposed mechanism avoids charge currents and thus does not cause strongly damping ohmic losses. Specific features that could further influence and verify the proposed alternative mechanisms of nerve conduction are discussed.

1. Introduction

Triggered by recent work presented Matthias Schneider in September 2018 at the NIM Conference "The Future of Nanoscience" as well the Article by Douglas Fox "The Brain, Reimagined" (1) in the April 2018 Issue of Scientific American I spent the last 10 months with some own basic experiments and literature research on the issue of how mechanical waves could efficiently transmit nerve signals. In this I used some insights gained through two decades of physics research on NanoElectroMechanical Systems (NEMS) studying signal transport in solid strings with nanoscale diameters. In addition, I utilized some basic notions about the propagation of interfacial waves on liquids that I experienced by surfing as well by observing the propagation of capillary waves in various containers. My way of thinking was also influenced by recent research of the role of water in controlling mechanical motion in biological systems, carried out by Peter Fratzl and collaborators (2, 3). To make the concepts discussed here accessible to a general readership interested in natural sciences ranging from physics and biochemistry to biology and medicine I will use only few equations and avoid derivations and instead use links to the knowledge base collected in Wikipedia and Google scholar as well as video demonstration available on YouTube to enable the readers to widen or fresh up their knowledge. I will make only few references to articles that are not easily accessible but instead try to list open access publications whenever possible. I hope that with my combination of existing knowledge a deepened understanding of potential electromechanical transport mechanisms of nerve signals can be achieved.

In a nutshell I will try to convince the reader that propagation of mechanical waves in tubular systems confined by elastic nanoscale membranes, filled and surrounded by biological water-based electrolytes is chosen by Nature to efficiently generate, transmit and convert neural impulses, so-called action potentials. In particular, I will discuss, why capillary waves generated by dipolar forces can propagate through the long tubular axons of neurons with small losses and transmit electrical signals with neutral dipolar currents. I will also comment why I consider this concept more realistic than the classical concept of the Hodgkin-Huxley model (4) based on propagation of electrical signals relying on ionic currents. In addition, I will describe, how ion injection induced by the activity of ion channels in the cell membrane of the soma will not only change the local electrical potential across the membrane but also induce capillary mechanical waves in the axon by changing the polarization of water close to the membrane. The resulting forces locally change the volume of the soma and, acting like a peristaltic pump, induce capillary mechanical waves along the axons accompanied by changes of the polarization of water dipoles near the Phospholipid membrane. of the axon. These polarization wave packets are of dielectric origin and can be considered to surf on the mechanical capillary wave along the axon. Their inherent strong electric fields can in turn induce chemical reactions and/or mechanical response in the synaptic termination of the neuron. This mechanism of coupled mechanical and dielectric waves is able to combine the roles of ion channels for locally modifying the polarization near the confining membrane with the one of mechanical waves to efficiently propagate the polarization change near the tubular walls of axons.

In my opinion previous models have neglected the potential role of rotating water dipoles that can cause strong mechanical forces in the presence of the large electric fields and their spatial gradients. Both exist in the narrow nanoscale region around the cell enclosing lipid membrane. Previous different electromechanical approaches that have tried to take the coupling of mechanical motion and electric signal propagation into account, have been recently reviewed by Drukarch et al. (5), expanded by further remarks by Holland et al. (6). However, none of them seem to consider the forces in confined liquids that are induced by the reorientation of water dipoles in the presence of strong electric fields which are the essence of the present approach.

In fact, with my collaborators we have discovered and studied a signal transport mechanism analogous to electromechanical surfing in quantum wells of piezoelectric semiconductors more than two decades ago (7). There we converted optical laser pulses into electron-hole pairs that, trapped in a moving sinusoidal electromechanical potential, were carried by a surface acoustic wave. Thus electron-hole pairs, spatially separated to suppress their radiative recombination, can surf over macroscopic distances and are controllably induced to reemit light at a location of choice without loss in photon energy. The underlying mechanism is sketched in the video Light surfing on SAW. It also relates to our more recent studies of coherent electromechanical wave packets propagating with little damping in insulating dielectric nanowires made out of silicon nitride (8, 9, 10). There electromechanical waves, coupling collective mechanical motion at interfaces with the modulation of dielectric properties, were utilized to generate and transmit signals with little loss along suitably designed nanoscale strings.

Starting with an introduction to capillary waves in elastic tubular systems such as neurons I will show that they will propagate with velocities that are in the same range as the measured velocities of the propagation of action potentials in axons. I will present arguments why such capillary waves are likely to be little damped and therefore suited to efficiently transport nervous signals. They thus exhibit essential properties that are required for propagating solitons, wave packets that preserve their shape over long distances, an important property of action potentials. I will then discuss why dipolar forces are well suited to induce electromechanical coupling, i. e. coupling of mechanical waves to electric fields and vice versa. I thus hope to present a consistent picture of how action potentials can be generated and propagated as neutral dipolar waves with little loss to transport and convert nervous signals efficiently. I will also suggest some experiments that could help to clarify whether the discussed mechanism presents a realistic description of the propagation of action potentials.

In the final chapter I will allow myself some rather speculative remarks on why Nature might have developed such a coherent wave mechanism for transmitting and storing our feelings and thoughts. I will speculate why some peculiarities in the construction of advanced neurons could have been introduced by Nature to improve their function. I also will include some potential links to other currently highly discussed branches of physics that may influence and perhaps improve our understanding of the function of nervous systems.

I hope that my remarks will generate strong interdisciplinary discussions and interactions to provide a deepened understanding about how we sense, transmit and combine neural signals. In the long run this may provide a better understanding of our nervous system and hopefully help in battling neural diseases.

2. Capillary waves in elastic tubular systems

The general structure of a neuron is displayed in Fig. 1. It basically consists of three parts, the cell body called soma that generates an action potential, the long tubular axon that transports the action potential from the soma to the synaptic terminals, and the synapse, which serves to route and possibly store the action potential via chemical signals.

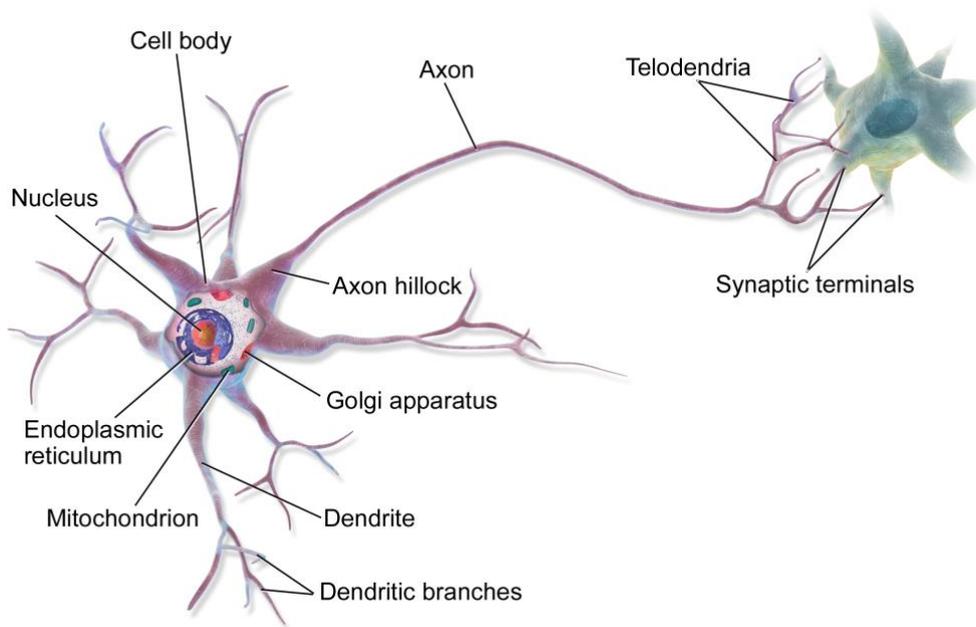

Fig. 1 Structure of a multipolar neuron
(https://upload.wikimedia.org/wikipedia/commons/1/10/Blausen_0657_MultipolarNeuron.png)

In the following I want to discuss, whether axons can be considered as quasi-one-dimensional waveguides for capillary liquid waves. I hope to convince the reader that it is likely that axons act in the similar fashion as the thread of a tin can telephone that my generation used as kids for local communication. In general, one may describe capillary waves as waves in liquids at an interface between two different media resulting from interface tension. The most common example are short wavelength waves at the surface of water that are generated for example by dropping a small object into a pond as seen in the video Ripples in a Pond. Whereas the excitations with long wavelengths **λ** on such a surface are dominated by gravitational forces and decrease in speed with decreasing wavelength (normal dispersion) the excitations with short wavelength **λ** are dominated by interface tension and increase their speed with decreasing wavelength (abnormal dispersion) as shown in Fig. 2.

The latter are dominated by capillary forces, e. g. at a water-air interface, and thus are called capillary waves. Oscillating with frequency f their angular frequency ω=2πf depends nonlinearly on wave vector **k**= 2π/**λ**, which is a measure of the momentum carried by the wave and increases with decreasing wavelength as shown in Equation (1).

**(1)**    $\omega^2 = \frac{\sigma}{\rho} |k^3|$

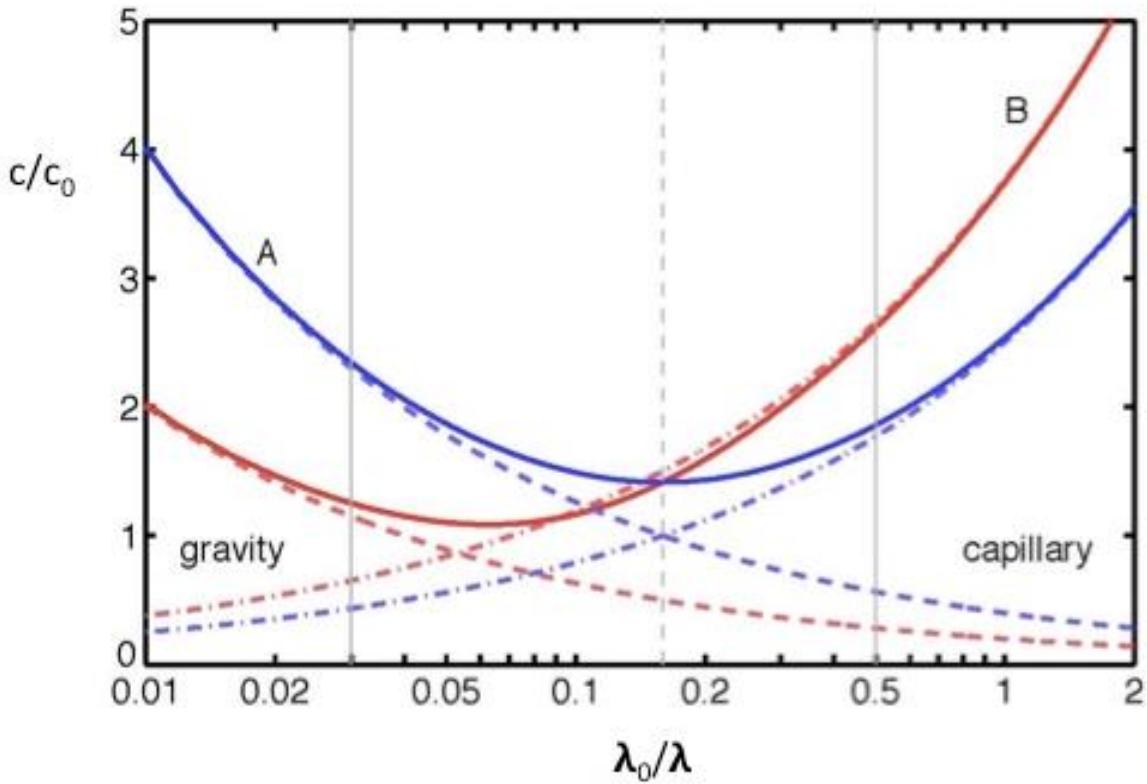

Fig. 2 shows the individual contributions of gravity forces (dashed) and capillary forces (dash-dotted) to the phase velocity $c_p/c_0$ (blue) and group velocity $c_g/c_0$ (red) of surface waves on water as a function of $\lambda_0/\lambda$ and their joint influence (solid) in normalized units with the gravity dominated regime A on the left and the capillary dominated one B on the right. The normalization constants $c_0$ and $\lambda_0$ are respectively, $c_0 \approx 23$ cm/s and $\lambda_0 \approx 0{,}27$ cm at room temperature and normal pressure (after https://upload.wikimedia.org/wikipedia/commons/9/9d/Dispersion_capillary.svg).

Here $\sigma$ denotes the surface tension at the interface and $\rho$ the mass density of the fluid (for water about 1g/cm$^3$). Hence one obtains for such capillary waves a phase velocity of

(2) $\quad c_p = \frac{\omega}{k} = f\lambda = \sqrt{\frac{\sigma}{\rho}k}$ and as group velocity of $c_g = \frac{d\omega}{dk} = \frac{3}{2}\sqrt{\frac{\sigma}{\rho}k}$ .

Another fact that supports the highly non-linear behavior of mechanical interface waves in confined liquids is long known as Plateau- Rayleigh instability. If we consider a cylindrical stream of water as ejected from a small tube with diameter D≤5mm we find that the originally cylindrical stream transforms into a stream of droplets as shown in Fig. 3a. This phenomenon reflects the non-linear dynamics of tubular water flow under the influence of capillary forces. It is driven by surface tension at the water-air interface trying to minimize surface area by the formation of droplets and can be viewed in the video Flow from a tap.

In a cylindrical elastic tube with a thin wall of thickness h<<D filled with and surrounded by the same liquid capillary forces on the inside and the outside of the tube will compensate. However, a positive pressure difference between the inside and the outside of the tube will stretch the tube wall and causes an effective circumferential interface tension acting along the tube perimeter. In response to a local pressure variation at the tube entrance these interfacial forces generate waves with both transverse and longitudinal motion in which the tube diameter oscillates periodically along the tube axis as pictured in Fig. 3b. For an incompressible fluid such as water the transverse oscillation of the tube diameter is coupled to oscillatory motion along the tube axis in order to keep the total volume of the incompressible liquid confined in a period section along the tube constant.

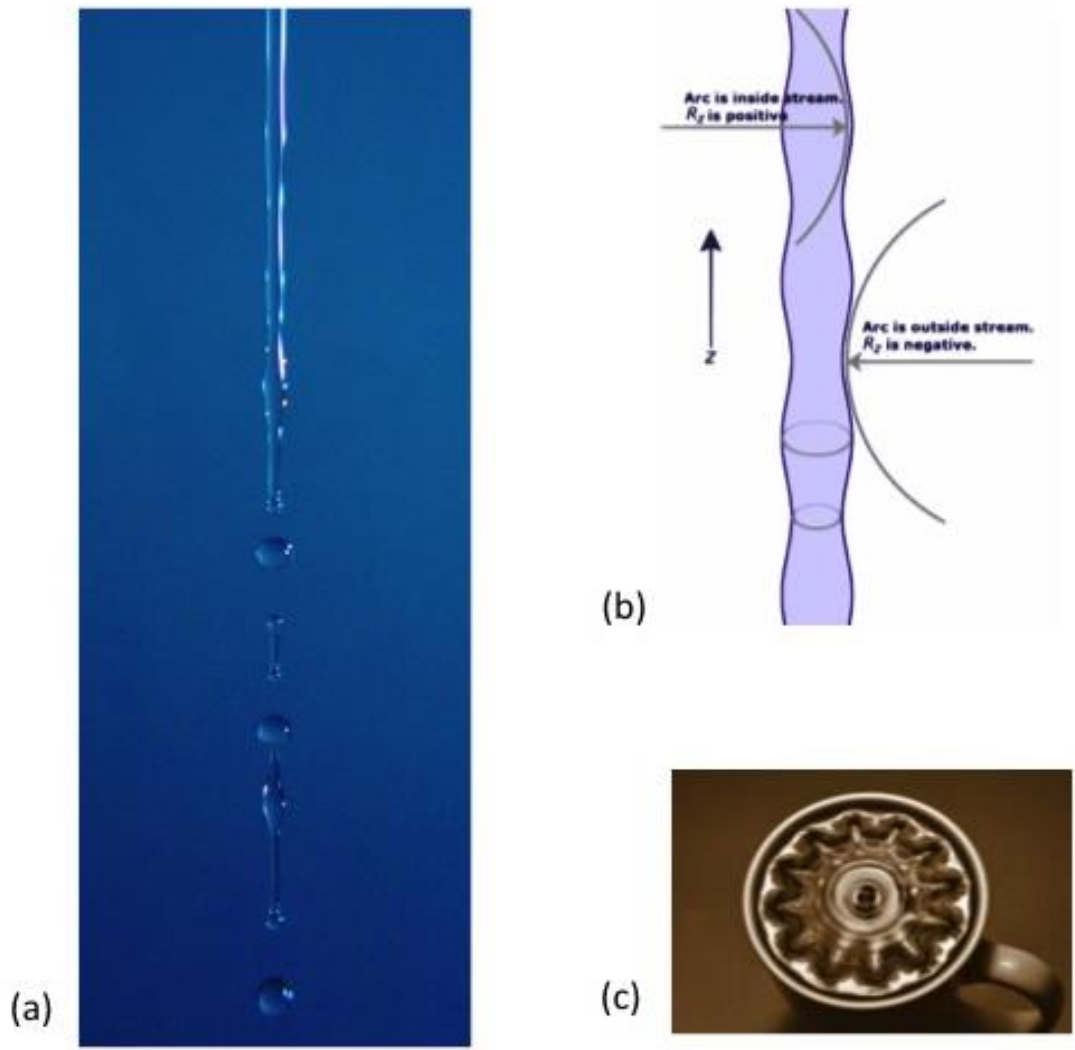

Fig. 3: (a)The Plateau-Rayleigh instability, the deformation of a free falling water stream with an initial diameter of about 3 mm by interface tension (From Ref. 7 of this link), (b) Modulation of the transverse mode diameter of an elastic water-filled tube by interface tension (c) Image of a higher order whispering gallery mode in a coffee cup. (From Normal modes, Wikipedia)

Whereas longitudinal pressure waves in rigid tubes propagate with the sound velocity, which in water is about 1500m/s, these interfacial waves usually travel with velocities well below the sound velocity as controlled by the effective elastic modulus E of the tube wall. The lowest propagating mode is a breathing mode in which the tube diameter D oscillates periodically in time with frequency f and spatially along the tube axis with wavelength $\lambda_L$ as indicated in Fig. 3b. This may be considered as a lowest frequency whispering gallery mode with half its transverse wavelength $\lambda_T$ equal to the tube diameter, i. e. $\lambda_T/2 \simeq D$ with axial mode index n=1 indicating that the motion has a zero-crossing in the center. In the cross-section of the tube this transverse mode (01*-mode) is ring-shaped with a oscillating radial amplitude near the tube wall, independent of the position along the perimeter, and zero amplitude on the tube axis. For illustration a photograph of such a whispering gallery mode in a rigid wall coffee cup with n = 6 transverse wavelengths fitting in the perimeter is shown in Fig. 3c and would have a mode index (0,6) reflecting that n=6 transverse wavelength fit in the perimeter, i. e. $6\lambda_T \simeq \pi D$.

In an elastic cylindrical tube with radius R filled with an incompressible liquid a short pressure pulse or a sinusoidal pressure variation at the frequency f of the lowest transverse elastic mode of the tube at one end locally modulates the tube radius R. It thus initiates a capillary wave oscillating at frequency f that propagates along the tube with speed $c_P$ and wavelength $\lambda_L$. Such tubular interfacial waves have been studied more than a century ago in order to understand the propagation of heart pulse signals along arteries of our body (11,12). In their simplest form of a fully elastic tube with free walls of thickness h<< R the resulting phase velocity, often called pulse phase velocity, is described by the Moens-Korteweg equation named after their discoverer in 1878.

**(3)** $\quad c_p = f\lambda_L \approx \sqrt{\frac{Eh}{2R\rho}}$

This is similar to the phase velocity of capillary waves at a plane water surface since the term Eh represents the incremental interface tension of the elastic tube wall and 2R is the tube diameter, which for the lowest transverse (01*)-mode is D≃$\lambda_T$/2. For a typical E of latex rubber of E ≈ 200 kPa , a wall thickness of h=0,25 mm and a tube diameter of D=5 mm one predicts $c_P$ ≈ 3m/s, roughly the same as found in own experiments discussed below. The influence of damping and of tethering by the environment is, e. g., described in detail in the 2011 review by Hodis and Zamir (12) and depends on the modeled boundary conditions. As discussed below for the lowest transverse mode these may not apply and since the damping observed in the experiment shown below is relatively low it will be neglected in most of the following discussion.

In a simple setup, usually employed to demonstrate wave propagation with surface waves of water I have studied the propagation of low frequency capillary waves in long rubber tubes with radii R ≲ 3 mm and wall of thickness h≲ 0,5 mm, freely suspended below the water surface in the tank as shown in Fig. 4. Exciting capillary waves at one end with periodic pressure variations of frequency f or short pressure pulse at one can observe the propagation of capillary waves along the tubes in space and time by watching the excitations of regular surfaces waves governed by the velocity dispersion in Figure 2. These are initiated by the motion of the tube walls through the propagation of a transverse mode capillary waves with frequency f in axial direction. For the tube with D ≈ 5 $mm$ shown in Fig.4 the lowest capillary wave mode has a resonance frequency f ≈ 6$Hz$ as measured with periodic pressure excitation. The oscillatory motion of the tube walls cannot be directly seen in this simple setup but its presence is detected by the excitation of the surface wave fronts at 6Hz, travelling nearly perpendicular to the rubber tube axis This is seen in video *_Tube5_*, whenever a pressure pulse launches a capillary wave packet. Illumination from above causes refraction of light on the wave fronts of the surface waves, seen as an intensity modulation of light shining at the screen below the tank after a reflection from a mirror tilted by 45 degree from the water surface. Note that the projection increases the image by a factor of 2, so that the 6cm long dark horizontal scale bar on the screen corresponds to 3cm at the water surface. The generation of the surface waves by short pressure pulses entering the tube at the left is shown in the video *Tube5* and are also seen in Fig. 5 which displays a still photograph extracted from *Tube5*.

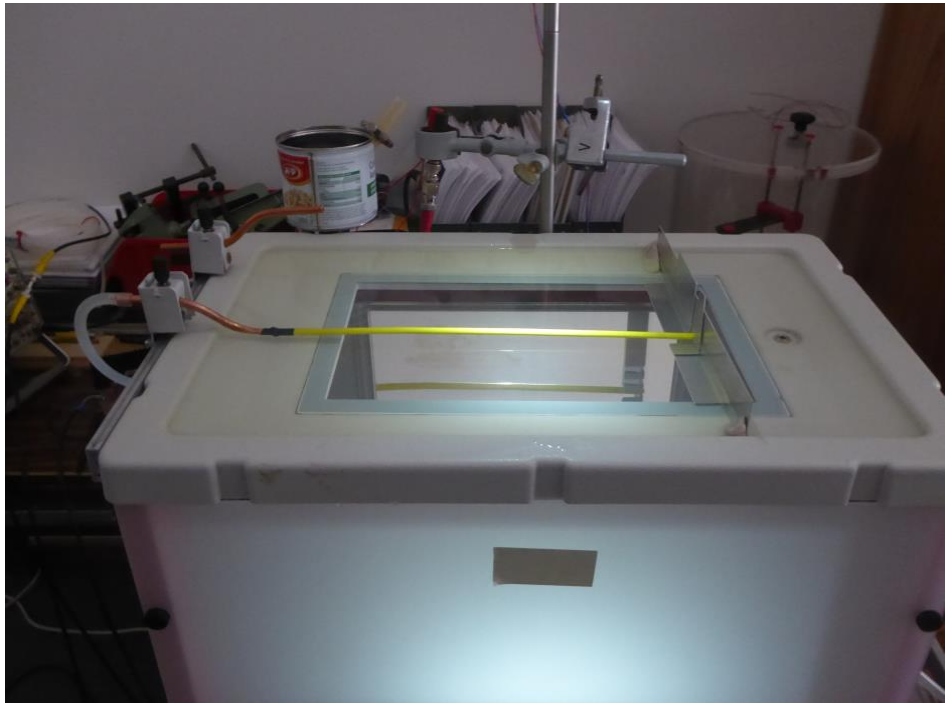

Fig.4: Experimental setup to study capillary waves propagating through a flexible rubber tube (yellow, diameter 2R ≈ 5 mm, wall thickness h ≈ 0,25 mm) filled with water at slight overpressure and suspended a few millimeters below the water surface. With short pressure pulses or sinusoidal pressure oscillations injected from the left capillary waves propagating in the tube excite waves on the water surface with wave fronts forming a small angle with respect to the tube axis. This reflects the significantly larger velocity of capillary waves in the tube in comparison to the ones on the water surface which propagate with the phase velocity c ≈ 1.5 $c_0$, determined by their wavelength **λ** (blue curve in Fig. 2). A LED spotlight above the water surface projects these surface waves via a 45 degree mirror mounted below the transparent tank onto the front screen with a 2-fold magnification, such that the marker of 6cm length on the screen corresponds to a length of 3cm on the water surface.

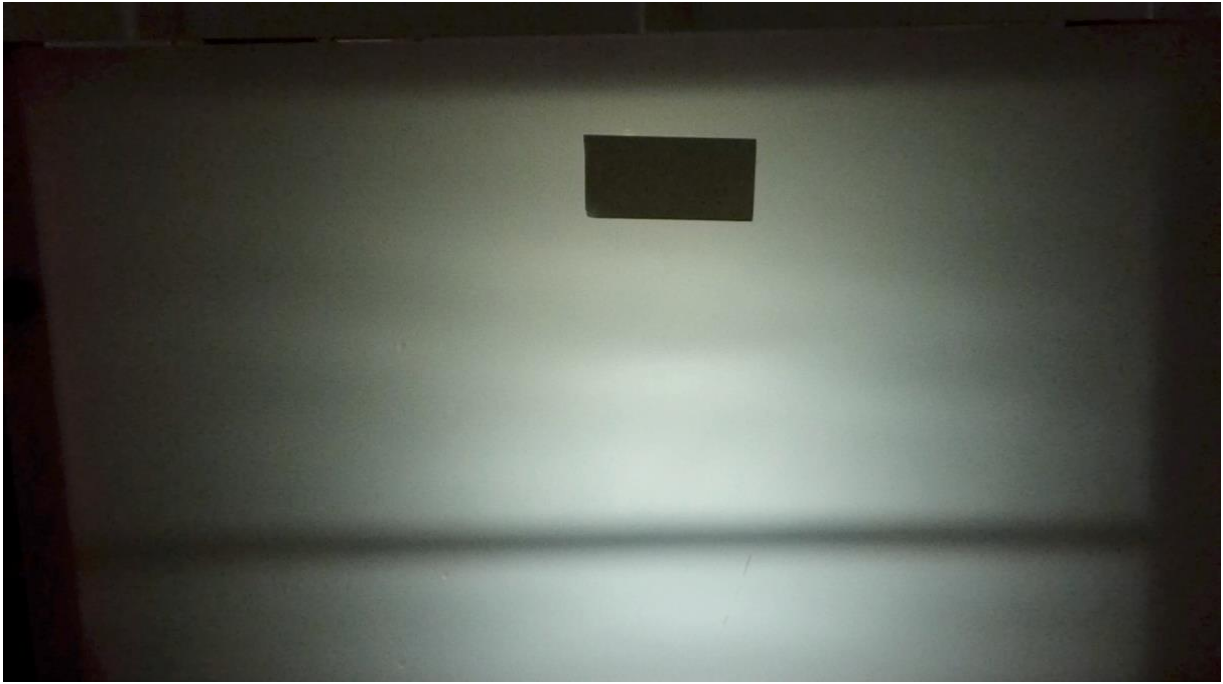

Fig. 5: Still photograph of the wave pattern of surface waves observed with the setup of Fig. 4 after a pressure pulse initiates a wave packet in the suspended rubber tube with diameter D ≈ 5 mm which in turn excites capillary waves on the water surface. The wave fronts of these surface waves with wavelength $\lambda/2$ ≈ 1.6 cm form an angle of about 10 degree with the tube axis, reflecting the fact that the waves in the tube are about 10-times faster than the surface wave propagating at c ≈ 1.5 $c_0$ . This value of $c_P$ ≈ 3.5 m/s agrees well with what is expected from Eq. 2 . The still is extracted from the video *Tube5*. The fact that the amplitudes of the surface waves do not decay appreciably from left to right reflects that the damping of the waves in the tube is comparatively small.

Two features are noteworthy. With the capillary waves in the tube travelling from left to right the wave fronts of the surface waves are oriented at an angle of about 10 degrees from the axis of the tube. This reflects that the velocity of the capillary waves in the tube is about a factor of 10 faster than the one of the capillary surface waves detected in the video Tube5. Using the dispersion shown in Fig. 2 with $c_0$ ≈ 0.23 m/s yields for the tube wave a velocity of about 3.5 m/s. Note that with Eq. 3 one predicts longitudinal wavelength $\lambda_L$ ≈ 60 cm, which can also be extracted in the experiments by seeing standing wave patterns if the rubber tube is terminated with a reflecting stop (not shown). The second observation in Fig. 5 is that the amplitudes of the generated surface wave do not change appreciably from left to right, reflecting that the amplitude of the capillary wave in the tube is not damped much by their excitation of such surface waves. This implies that the coupling between the two types of waves is relatively small. This is not surprising since the two phase velocities Eq. 2 and Eq. 3 differ substantially. In conclusion, we learn that the velocity of the capillary waves in the tube can be well understood with the classical Moens-Korteweg equation neglecting any damping by the environment. Hence the capillary waves in the tube propagate with relatively little loss, as generally may be expected in a one-dimensional waveguide with little coupling to the environment as discussed below.

With this understanding one can predict from Eq. 3 that for tubes with diameter of $10 \mu$m filled with a incompressible liquid of the density of water and an interface tension of about 10-100 mN/m, typical for lipid membranes encasing axons, phase velocities of about 1-3 m/s can be expected, in good agreement with typical measured values of action potentials in axons that are not stiffened by myelination. Effective stiffening of the lipid membranes by a myelin sheath can most likely increase these values by up to 10-fold, in rough agreement with measured values of the increased velocities of action potentials in myelinated axons shown for comparison in Fig. 6 and in Table 1. In contrast to the traditional Hodgkin-Huxley model no electric network with high losses that slows the velocity of electromagnetic signals from 300 000 km/s to about 30 m/s is needed to explain the observed propagation velocities of action potentials but they can be directly predicted by Eq. 3. Furthermore, the coherence of capillary wave packets can be expected to be high, so that the action potential can travel quasi adiabatically and soliton-like without changing its characteristic signature. In addition, since Nature has selected propagation of capillary waves at liquid-membrane interfaces confined in quasi-one-dimensional tubular geometry to efficiently transport the heart pulse signal through arteries, I consider it highly unlikely that Nature has independently developed a completely different and rather lossy mechanism to propagate action potentials by ion currents as described by the Hodgkin-Huxley model.

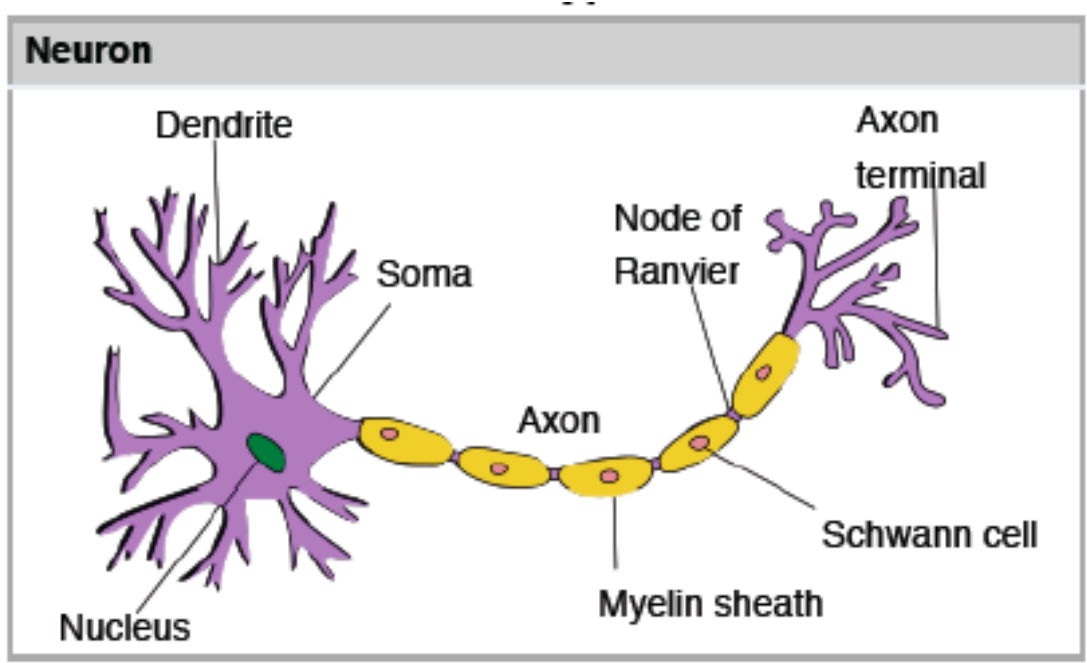

Fig.6: In a Myelinated Neuron with a stiff myelin sheath surrounding the axon between the nodes of Ranvier one observes an about 10-fold increase of the propagation velocities of action potentials in comparison to unmyelinated axons (from Wikipedia).

### Motor fiber types

| Type | Erlanger-Gasser Classification | Diameter | Myelin | Conduction velocity | Associated muscle fibers |
|---|---|---|---|---|---|
| α | Aα | 13–20 μm | Yes | 80–120 m/s | Extrafusal muscle fibers |
| γ | Aγ | 5–8 μm | Yes | 4–24 m/s [2][3] | Intrafusal muscle fibers |

Different sensory receptors are innervated by different types of nerve fibers. Proprioceptors are innervated by type Ia, Ib and II sen by type II and III sensory fibers, and nociceptors and thermoreceptors by type III and IV sensory fibers.

### Sensory fiber types

| Type | Erlanger-Gasser Classification | Diameter | Myelin | Conduction velocity | Associated sensory receptors |
|---|---|---|---|---|---|
| Ia | Aα | 13–20 μm | Yes | 80–120 m/s[4] | Responsible for proprioception |
| Ib | Aα | 13–20 μm | Yes | 80–120 m/s | Golgi tendon organ |
| II | Aβ | 6–12 μm | Yes | 33–75 m/s | Secondary receptors of muscle spindle<br>All cutaneous mechanoreceptors |
| III | Aδ | 1–5 μm | Thin | 3–30 m/s | Free nerve endings of touch and pressure<br>Nociceptors of neospinothalamic tract<br>Cold thermoreceptors |
| IV | C | 0.2–1.5 μm | No | 0.5–2.0 m/s | Nociceptors of paleospinothalamic tract<br>Warmth receptors |

Table 1: Nerve conduction velocity in neurons (from Wikipedia)

3. Some remarks on the damping of capillary waves in tubular systems

Before discussing the electromechanical generation and character of action potentials in neurons let us consider possible damping mechanisms for capillary waves serving as carriers of action potentials. As the tube confines the water inside and water is an incompressible liquid a local change of the tube radius acts in a similar way as a peristaltic pump and will generate a capillary wave pulse along the tube. At the small tube radii considered here liquid flow is usually laminar and we can neglect any effects of damping by turbulence. When we consider the propagation of capillary waves with transverse wavelengths $\lambda_T$ corresponding to twice the tube diameter they will propagate in the tubular waveguide in the regime of the lowest transverse mode. In analogy to light in optical fibers or microwave radiation in hollow metallic waveguides we can consider the propagation of capillary waves guided by the tube wall as quasi-one-dimensional. Without additional irregularities in the tube diameter or unwanted motion of the tube wall by environmental phenomena on lengths scales comparable to the longitudinal wavelength $\lambda_L$ one can expect the capillary waves to propagate nearly undamped over relatively large distances. They are hindered by momentum conservation laws from losing their energy and momentum to other low energy modes of motion. With their large longitudinal wavelengths $\lambda_L$ their momentum $k=2\pi/\lambda_L$ is small in comparison to other collective interface excitations. Accordingly, their phase velocity is in a range of values well above other collective mechanical interface excitations but at the same time well below longitudinal pressure waves propagating with sound velocity in biological media. This decreases their coupling to the environment and is probably why one observes only little damping of the capillary tube waves as in the water tank experiments shown above. The fact that elastic backscattering of the waves, i. e., a reversal of their velocity is unlikely, results from the quasi-one-dimensional guidance in the tube. Qualitatively the same applies for higher order transverse whispering gallery modes with index (0n) with n >1 but still small such that neighboring transverse modes remain energetically distinct and thus oscillate at sufficiently different frequencies f.

Exploring the resonant behavior of mechanical waves in solid strings of nanoscale diameters in vacuum, which also act as low mode quasi-one-dimensional waveguides, we found in recent studies of coherence and dissipation that energy loss occurs essentially only at the end of the strings, where they are clamped by the support and elastic scattering causing decoherence is even more unlikely (9,10) Therefore one can easily achieve in nano-strings suspended in vacuum quality factors Q = $\omega/\Delta\omega$ of a million. This is a reciprocal measure of dissipation as it divides the resonance frequency $\omega$ of a string resonator by the rate of dissipation $\Delta\omega$ and is a measure of coherence. By tailoring the clamps of the string to efficiently reflect the energy of the mechanical waves a group in Lausanne (13) recently achieved record Q-factors of such nano-strings at room temperature of close to a billion. This shows convincingly that energy loss of mechanical waves in solid nanoscale strings propagating in a low transverse mode can be dramatically reduced so that wave transport is quasi-adiabatic. High Q narrow-band electromechanical filters in high frequency communication devices such as mobile phones are using since decades mechanical filters based on acoustic waves in piezoelectric chips to separate frequencies and assure that we do not have to listen to all other mobile phone users in our vicinity. In comparison, optimized electronic filters usually can only achieve Q-factors that are in the range of a few hundred. For the same basic reason our quartz watches use mechanical oscillations to minimize energy loss and thus achieve high precision timing.

The fundamental difference between the propagation of electric signals along metal wires and mechanical waves propagating in low mode waveguides is that electrons or ions as small entities can usually be easily scattered by the presence of irregular positioned fixed charges. Such quasi-elastic scattering releases momentum conservation rules and dominates decoherence and the Ohmic resistance. In contrast, mechanical modes represent a collective motion of many strongly coupled atoms or molecules where scattering by small scale local disturbances can be effectively suppressed. This one can observe in many musical instruments by listening to the sound of a stretched string for many seconds (with a Q ≳ 1000) or the propagation of mechanical pulses in the well-known Newton's cradle. This common toy consists of a chain of pendula of string-suspended steel balls aligned in a quasi-one-dimensional straight file such that they touch each other. When the pendulum at one end is put into motion its energy is transferred nearly without loss to the pendulum at the other end of the chain as a consequence of momentum conservation and reverses its direction with each swing for many oscillations. In general, one finds that collective wave modes in strongly coupled matter such as water or chains of pendula can propagate rather undisturbed over long distance when the flow is laminar. This is also convincingly reflected by the observation that an earthquake in Chile that initiates collective ocean waves can cause a Tsunami in Hawaii. This demonstrates dramatically the preservation of coherent interference of water waves over huge distances. Also ultrasound attenuation in viscoelastic biological materials is small and scales roughly linear with frequency f and length, exhibiting typical values of 0.5dB/(MHz . cm). This means that at a frequency of 100 KHz the length at which the intensity of an ultrasound wave decays by a factor of two is about 60 cm. This is why we can image all parts of our body with ultrasound.

With the arguments used above one can expect that lateral confinement of such mechanical waves in tubes may further decrease damping. Insofar it is not surprising that in studies of pressure pulse induced propagation at an water-lipid-membrane interfaces hardly any damping was observed over distances of 20 cm corresponding to attenuation coefficients of order 1/m (14). This might even be further reduced in a quasi-one-dimensional geometry as in our study of capillary waves in narrow tubes, where the only elastic momentum scattering process is backscattering, i. e., a complete reversal of momentum. Such highly coherent mechanical signals also cause the pulse wave signals observed in our arteries to not substantially change their shape with increasing distance from the heart. Elastic backscattering of capillary waves in narrow tubes operating with a discrete low transverse mode does not change the shape of the pulse but only decreases its forward moving amplitude. Because of the above arguments I consider the propagation of action potentials as a highly coherent non-equilibrium process. This view is also supported in the next section by a video showing the propagation of channeled surface waves of water behaving as solitons.

4. The role of solitons in the transmission of action potentials.

In previous publications by Heimburg et al. it was proposed that action potentials are mechanical solitons propagating at membrane-liquid interfaces (15,16). Hence one can ask what role solitons play in wave propagation in confined liquids. Wikipedia explains a soliton as a "self-reinforcing solitary wave packet that maintains its shape while it propagates at a constant velocity. Solitons are caused by a cancellation of nonlinear and dispersive effects in the medium".

The fact that solitons can propagate in water channels and that heart pulse waves as discussed above propagate at constant speed without changing their shape makes it likely that solitons can exist in capillary wave transport. Capillary waves are dispersive and change their velocity with inverse wavelength (as seen in Fig. 2) and are also nonlinear in amplitude. Such nonlinearities are further enhanced if the confining membrane approaches a structural phase transition (17,18,19). Solitonic shallow water waves can be observed even in macroscopic channels as seen in Water channel solitons and the dynamics of solitons is also well illustrated in the video Visualizing solitons., e. g. with coupled pendulums. As illustrated in Fig 7 such a soliton wave packet can also consist of a higher carrier frequency (blue) and an envelope (red) that does not change during propagation.

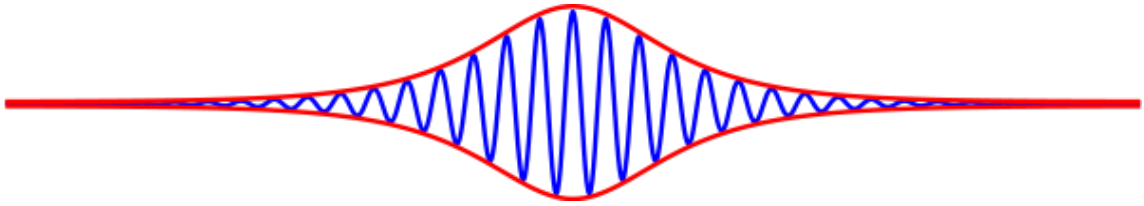

Fig. 7: A hyperbolic secant (sech) envelope soliton for water waves:
The blue line is the carrier signal, while the red line is the envelope soliton.
(From Wikipedia)https://en.wikipedia.org/wiki/Soliton#/media/File:Sech_soliton.svg

In the next section I will try to show that water molecules, coupled by dipolar interaction, can be collectively rotated in their orientation with respect to an interface through changes in the electric landscape. Such polarization wave packets defined by the locally changing dipole orientation can propagate through an axon as a tubular capillary wave. Such a view is also supported by the detection of mechanical motion previously observed along with action potentials in axons as well as in planar cellular membranes as studied by Wilke, Tasaki, Takaki, Heimburg, and Schneider (5).

5. Electromechanics with dipolar forces

Action potentials in nerves are excited by local changes of the electrostatic environment and often caused by the opening and closing of ion channels in the cell membrane. With the so-called patch clamp technique developed by Bert Sakmann and Erwin Neher (Nobel Laureates in Medicine, 1991) one can study the reaction of ion channel currents to chemical and physical changes of the cell environment. This is a widely used technique to study the influence of pharmaceutical drugs onto action potentials and today is performed highly in parallel in patch-clamp robots, fabricated, e. g., by Nanion Technologies, a spin-off of our interdisciplinary research at CeNS. The role of ion-channels has led to explaining the propagation of action potentials by the Hodgkin-Huxley-model based on ion currents. One drawback of this model is that one has to use a rather complex description by electric cable theory to reduce the velocity of electromagnetic signals from its vacuum value of 300.000 km/s to the measured values of action potentials in the range of 0.1 to 100 m/s. Another drawback is that Ohmic losses induced by diffusive ion currents through large resistances have to be compensated by complex amplification processes in order to enable the propagation of action potentials of constant shape and amplitude over distances up to several meters. Now that we have learned that mechanical capillary waves can propagate soliton-like with low loss through narrow tubes serving as waveguides one has to consider how such wave packets influence the electric environment and vice versa how the electric environment influences the propagation of capillary waves. Here we can profit again from knowledge gained via recent research on Nano-Electro-Mechanic Systems (NEMS).

With my collaborators we introduced a simple and effective way to excite transverse oscillatory modes of nanomechanical strings, made out of insulating Silicon nitride by dipolar forces (8). Basically, this uses the effect of gradient fields on dielectric materials in a way similar as employed in optical tweezers, an invention for which Arthur Ashkin was awarded the 2018 Nobel Prize in Physics. We use a local change of a large electric field generated by two closely spaced electrodes in close vicinity to an insulating nano-string as shown in the electron micrograph in the left of Fig. 8 to electromechanically excite and also sense high frequency transverse mechanical oscillations in such nano-strings. Using electrode spacings in the submicrometer range one can easily generate with low voltages in the range of a few Volts static electric fields with strengths of typically $10^7$ V/m in the vicinity of electrodes and thus polarize the dielectric strings along the field direction perpendicular to the string axis as illustrated in Fig. 8. This effectively induces an electric polarization of the string pictured as an electric dipole. This is attracted by a force parallel to the gradient of the electric field as dipoles are "high field seekers".

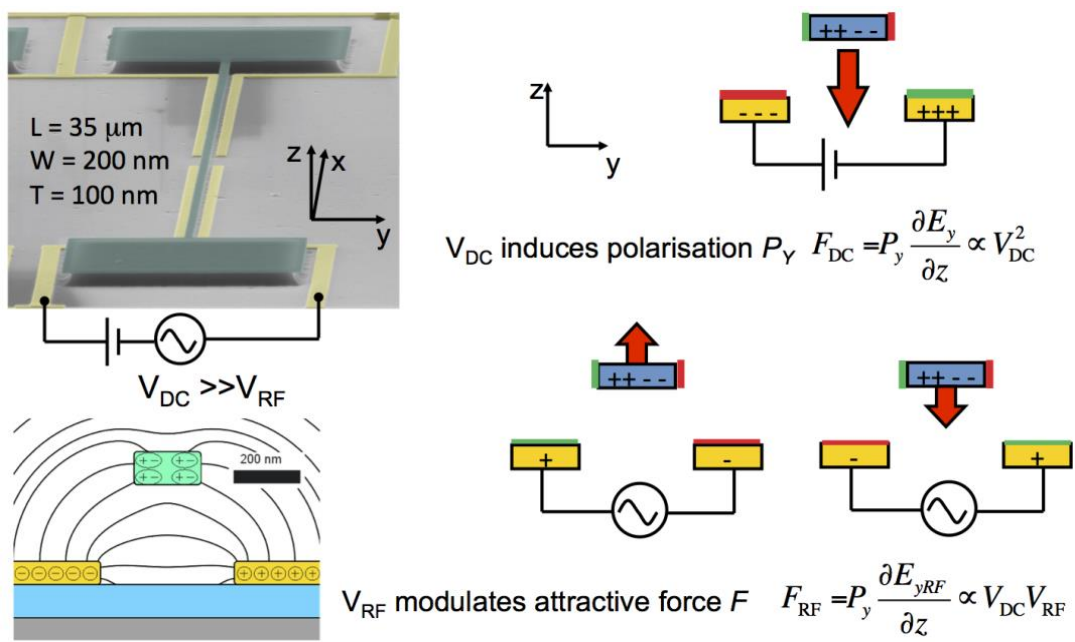

Fig. 8: Illustration of the mechanisms of dielectric excitation of a stretched nano-string of insulating silicon nitride (green) suspended above a pair of electrodes (yellow). A static voltage $V_{DC}$ between the electrodes induces a polarization $P_y$ in the y-direction of the string. The electric field gradient $dE_y/dz$, discernible by the uneven spacing of the field lines indicated in the cross-sectional view (lower left), causes a force $F_{DC}$ that attracts the string towards the electrodes. An additional small voltage $V_{RF}$ oscillating at radio frequencies $dE_{yRF}/dz$ modulates this force and can excite transverse mechanical oscillations of the string at its characteristic resonant frequencies. Thus one can electromechanically play a nano-guitar. The second set of electrodes can be used to capacitively detect the string motion Thus one can electromechanically play and listen to a nano-guitar /after 8 /.

In my opinion the role of such dielectric forces has not been taken sufficiently into account in the traditional Hodgkin-Huxley model nor to my knowledge in any of the present models of mechanical propagation of action potentials in neurons. It deserves a more detailed attention since water, the main constituent of cellular electrolytes, consists of molecules with large electric dipole moments. Their orientation in external electrostatic fields causes its large electric susceptibility of about $\chi \approx 80$, describing the response of a polarization to an electric field. In bulk water dipoles can freely rotate and will effectively align in a large external electric field in order to screen, i. e. weaken the external electric field. Contrary to ionic charges governed by Coulomb forces (which repel each other when of equal charge and attract each other when of opposite charge) electrically neutral free dipoles will always be attracted to the region of strongest electric field because of gradient forces, following the same rules as the dielectric string in Fig. 8.

In a cellular environment, the highest electric field strength usually exists across the insulating lipid cell membrane and a narrow water layer between the electrolytic cell interior and the insulating cell membrane.  There the electric field is confined to a layer of typically 10 nm thickness, consisting of the usually insulating lipid cell membrane  (5-7 nm thick) and the adjacent Debye screening regions in the liquid electrolyte (typically 1nm each) containing both ions and predominantly water dipoles. Recent theoretical and experimental studies have indicated that at polar interfaces water dipoles will orient themselves perpendicular to the interface. This is because the electric environment at the interface depends on effective local fields reflecting the nature of composition-dependent dipolar forces between the lipid membrane and the water molecules in the adjacent electrolyte. These studies have let to the so-called flip-flop model ([20]) in which the two opposite dipole orientations can lead to bi-stable configurations, in analogy to electronic flip-flop circuits. Depending on the strength and the direction of the local interface electric field and the hydrophobic or hydrophilic nature of the membrane interface the water dipoles of the first few molecular layers adjacent to the membrane can exhibit nearly perfect orientation perpendicular to the membrane interface similar as dipoles in ferroelectric (or antiferroelectric) materials ([21]). This is demonstrated in optical studies as well as in simulations (see,e. g.  [22],[23]). Some authors classify this rather stable polarization state as "ice" of oriented dipoles ([24]). I rather picture that the strength and orientation of the electric field perpendicular to the lipid membrane can induce quasi-ferroelectric domains of highly oriented dipoles that can flip under a reversal of the electric field direction. Note that transmembrane voltages in neurons usually do not exceed 100mV so that electrolytic dissociation of the water dipole is unlikely to occur in spite of the application of high electric fields.

Such a view gains also support from recent studies of the behavior of water confined in narrow channels of varying width engineered with atomic layer precision by the Nobel Laureates Andrei Geim and Kostya Novoselov and collaborators ([25], [26]). With decreasing width it is found that the electric susceptibility measured in capacitance studies decreases from the high bulk value of water ($\approx$80) to only 1 for 1nm wide channels ([25]) in general agreement with theoretical predictions ([27]). This implies that small oscillating electric fields cannot significantly change the polarization of the oriented water dipoles in narrow channels, as expected for ferroelectric (or antiferro-electric) behavior in strong interface electric fields.

Since in neurons the electric field across lipid bilayer membranes is high and can be reversed with the voltage bias one might expect a hysteretic behavior of the dipolar orientation. Such a behavior is likely to have a threshold for the onset of collective reorientation and could exhibit Bistability (28). In the following I will therefore assume that the fields generated by action potentials are sufficiently strong to induce stable quasi-ferroelectric domains consisting of water molecules oriented perpendicular to the lipid membrane of the axon in the high field region. These polarized but charge neutral domains can surf on the capillary waves propagating through the axon. Gradient fields along the axial boundaries of such domains with opposing polarization induced by the local change of the axon diameter will move with the capillary wave and, inducing a flip of the water dipoles, will enable the propagation of such high field domains. Additional confinement of these dipolar domains along the propagation direction may result from the curvature in the axon membrane that propagates with the capillary waves and can further increase the local electric field strength at membrane. Figs. 9 and 10 try to illustrate how the action potential generated by ion injection into the soma modifies both the polarization of the water dipoles close to the cell membrane as well as the diameter of the soma and axon and thus induces a motion of the ferroelectric-like dipolar domains along the axon.

For simplicity let us consider the cell body, the soma of a neuron and its attached dendrites, as a closed container that only has an opening at the hillock of the axon. The electromechanical activity of the neuron will be dominated by the roughly 10 nm thick high electric field region around the insolating lipid membrane containing positive and negatively charged ions as well as water dipoles in the Debye screening layer partially oriented by local electric field. With the resting potential of -70 mV (left region in Fig. 9) the electric field of about $10_7$ V/m will cause an accumulation of aligned water dipoles at the interior of the lipid cell membrane. There the dipoles will be polarized with their negative oxygen pointing towards the effectively positive counter ions on the outside of the neuron as sketched in the lower left of Fig 9. They will not only decrease the effective local electric field but, attracted by the positive charge outside the membrane, also push the cell wall towards the outside environment. With an elastic membrane this force will expand the membrane and be compensated by the additional tensional force of the membrane to establish equilibrium.

Since the injection of positive ions into the soma decreases the cellular potential the polarization of the water dipoles diminishes while simultaneously the soma volume will shrink and push some positive ions as well as oriented water dipoles with decreasing polarization across the hillock into the adjacent portion of the axon as schematically depicted in Fig.10. This peristaltic pumping will launch a capillary wave packet enclosed by a skin of partially oriented water dipoles into the axon. In addition, the launching of the capillary wave into the axon by the volume decrease of the soma that accompanies depolarization will also lead to a presumably small

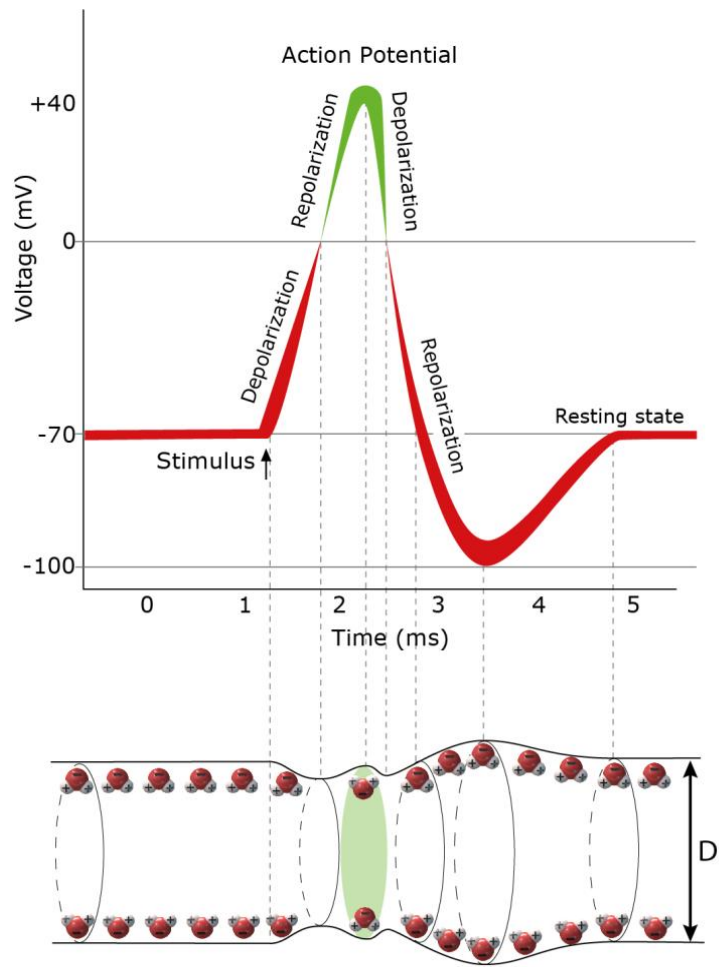

Fig.9: A typical action potential as measured by the temporal voltage change across the cell membrane and generated by a change of ion concentration in the soma launches a wave packet of modified polarization of water dipoles that, if observed at a fixed position along the axon is associated with a temporal change of the dipole orientation near the wall of the lipid membrane and its associated change of axon diameter D as indicated. The width of the line picturing the action potential qualitatively indicates the degree of polarization and is red (green) when the negative (positive) end of the water dipole faces the membrane. It neglects the possible occurrence of a phase transition from a paraelectric to a ferroelectric state at strong fields. The dipolar forces of the polarized water layer increase the axon diameter D at finite electric field strength. Note that the layer of polarized dipoles with a width of about 1nm is typically 10.000 times smaller than the axon diameter of roughly 0.01mm. (Graphics by Chris Hohmann, LMU)

pressure increase in the axon as observed already in early experimental studies of Terakawa (29). Fig. 9 schematically relates the typical temporal shape of the action potential at a fixed position along the axon to the change of the orientation of water dipoles and the accompanying change in the axon diameter D. Fig.10 reflects the spatial shape of the wave packet at an arbitrary time delay and thus has the inverse axial shape. Since the dipolar forces modulate the size of the axon diameter the whispering-gallery-mode capillary wave packet has a donut-like radial shape.

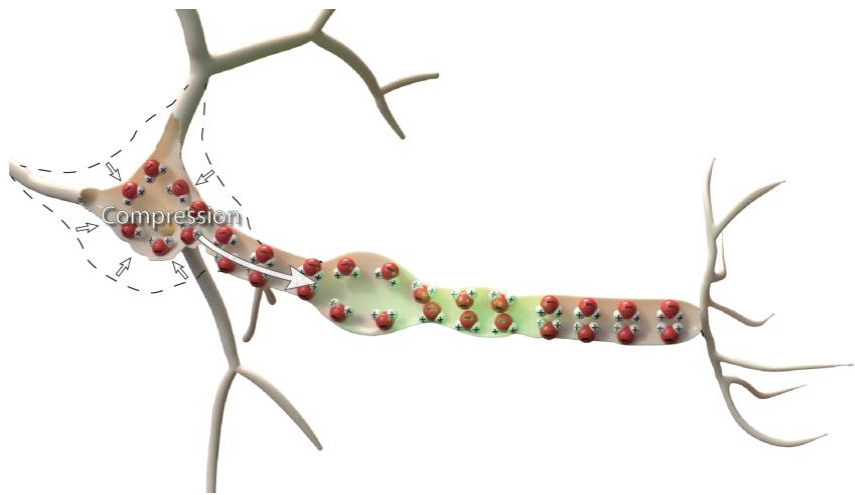

Fig. 10: Sketch of a neuron to illustrate how a compression of the soma caused by the injection of positively charged ions generates a moving wave packet of a capillary wave in the axon that carries a domain of changed polarization in the narrow electric field layer facing the tubular lipid membrane generated through the reorientation of water dipoles which is accompanied by corresponding change in the axon diameter. Here the domain of changed polarization is pictured in space at a fixed time and thus looks reversed to Fig. 9, where the temporal change of the wave packet is shown at a fixed position. Note that again the scales are grossly distorted. The length of an axon can be in the range of 1m and thus 100.000 times longer than the axon diameter and about as long as the size of the surrounding living being. (Graphics by Chris Hohmann, LMU)

Assuming that the axon does not contain many ion channels the positive charge on outside the axon wall beyond will stop the propagation of positive ions from the soma beyond the hillock. However, once the field-induced reorientation of the neutral water dipoles is stabilized by inter-dipolar interactions these polarization domains can continue to propagate trapped within the capillary waves with a velocity determined by the geometry and effective tension of the axon wall. A similar suppression of ionic motion but unhindered propagation of water molecules has now been observed in atomic size capillaries created by the Geim et al. group (26). As the action potential rises from the rest potential the decreasing polarization of the water dipoles in the high electric field regime around the axon diameter will decrease until the effective voltage across the axon wall drops to zero as indicated in Fig.9. At vanishing electric field across the axon membrane its diameter beyond the hillock will shrink to a minimum. This decrease in net polarization near the lipid membrane might be accompanied by a transition from a ferroelectric phase to an antiferroelectric phase since the dipolar interactions between the water molecules continue to be strong. As the effective potential at the hillock becomes more positive the water dipoles rotate to oppose the changed direction of the electric field and will again built up a polarization but now with opposite polarity. Hence the dipoles will again increase their push towards the lipid membrane of the axon as they again are attracted by the effective local electric field. Note that the change in polarization continuously increase until the effective action potential reaches its maximum at + 40 mV. Then the charge imbalance in the soma as well as in the entrance of the axon will reverse its direction and again the soma volume will decrease until the effective potential in the soma is zero. As the action potential becomes again negative the polarization of the water dipoles will built up again with opposite polarity and increase the volume of the soma and the diameter at the axon entrance. Correspondingly, there will develop an excess negative charge in the soma raising the effective voltage beyond the rest value to -100mV where the soma volume and the axon diameter at its entrance will be at a maximum. This overshoot, probably caused by the delayed response of the ion injection to the local electric field variation, will be reduced until again the rest potential and the corresponding soma volume are restored. In this process the maximum field strength across the lipid cell membrane is getting close but stays below the field for electric breakdown. Also remember that the curvature of the membrane wall in the axial direction around the extrema of the field strength will increase the local field strength and thus confine the polarization state of the water molecules within the extrema of the capillary wave travelling along the axon.

After the local variations of ion concentrations in the soma and at the entrance of the axon have launched the capillary wave packet impressed with the polarization signature of the action potential and the transverse pattern of the excited whispering gallery mode it will continue to propagate with the accompanying polarization wave along the axon. With no effective mechanisms for backscattering in this quasi-one-dimensional tubular mode of the combined electromechanical polarization wave this action potential pulse will not be damped and propagate until it can release its energy in the synapsis. A possible mechanism for this release could be an electrochemical reaction in the synapsis induced by the transient electric field of the action potential or a transfer to other nerves across the synaptic node.

To experimentally verify the described mechanism of charge-neutral dipolar capillary waves one needs to correlate the measured action potential with mechanical changes of both the soma volume and the axon diameter and show that both become minimal, when the effective local polarization of the water dipoles vanishes at a local transmembrane voltage of 0V. Whereas the local axon diameter will be minimal at this membrane voltage the moving polarization induced by the orientation of the water dipoles will change its sign whenever it crosses this voltage. Furthermore, one can expect that the propagation of the action potential and its accompanying dipolar polarization wave is partially reflected whenever a strong local modification of the axon diameter, e. g. caused by structural modifications such as the nodes of Ranvier, characteristic for myelinated axons pictured in Fig 6, induces a strong local electric field gradient along the axon lipid membrane.

The time constant that governs the dynamics of the thus described action potential probably reflects an [RC time constant](RC time constant) $\tau$ =RC, determined by the local capacitance change C and the local resistance R for the trans-membrane ion current and is likely to be longer than the period 1/f of the mechanical oscillation of the cell membrane depending on mode index n. Thus it can reflect a soliton-like excitation with an envelope that mirrors the action potential but can contain carrier frequencies that are higher than $1/\tau$, as indicated in Fig. 7. But they are not damped by a dissipative RC-time-dominated readout scheme which rather measures the envelope of the solitonic wave packet. Such an electro–mechanical scheme not only uses the collective nature of mechanical waves to carry the neutral changes in polarization but can also preserve the shape of the action potential as a soliton-like wave packet of modified polarization near the tubular wall of the axon. Note that the soliton-like character can only be realized by coherent, i. e. essentially undamped wave packets and cannot be realized in a highly dissipative system of diffusively moving ionic charges. Closely packed water dipoles that can be rotated by electric fields around the one-dimensional axis of capillary wave propagation and are coupled by dipolar interactions are conceptually very similar to pendula coupled by torsional forces that are evenly spaced on a string. The latter are often employed to demonstrate soliton-like wave packets. Insofar with little damping and suitably strong dipolar coupling it is conceivable that excitations of one-dimensional wave packets of water dipoles can exhibit soliton-like character.

The induced dipolar quasi-one-dimensional polarization wave packet and its accompanying mechanical deformation of the axon tube diameter is charge-neutral and results predominantly from a driven rotation of the dipolar water molecules without appreciable molecular motion along the direction of the moving wave thus causing negligibly small dissipation. In that sense the polarization wave surfs on the mechanical wave without appreciable movement of the water molecules in the direction of propagation.

Since in axons the polarization skin is part of the propagating capillary wave I am confident that neutral dipolar currents can be propagated by capillary waves of sufficiently high amplitude. Therefore surfing of wave packets of partially oriented water dipoles on coherent capillary waves is in my opinion a better model to explain the nature of action potentials than the Hodgkin-Huxley model based on dissipative ionic currents.

Since the velocity of the mechanical capillary wave is determined by the membrane stiffness it immediately explains both the general magnitude of the measured propagation velocities of the action potentials. It also explains why myelination of axons serves to achieve higher velocities for action potentials since it is most likely stiffening the axon membrane. This replaces the somewhat mysterious explanation of [saltatory conduction](#) with a much simpler explanation. As depicted in Fig. 6 myelinated axons consist of a chain of Schwann cells surrounded by a stiff myelin sheath periodically interrupted by nodes of Ranvier. In the model of [saltatory conduction](#) these nodes serve to inject ions to increase the driving longitudinal electric fields such that the action potential "jumps" from node to node. I find it hard to believe that these jumps are compatible with the dissipative nature of ionic currents. Instead, I consider it more likely that the stiffening of the axon tubes through the myelin sheath is responsible for the higher propagation velocities of myelinated axons, in general agreement with Eq. 3. In the more speculative final section I will present potential additional motives why Nature has chosen the more complex construction of myelinated axons to enhance propagation of action potentials.

With suitable excitation and detection electrodes and accordingly sensitive instrumentation one should be able to verify the here presented electromechanical model of the propagation of action potentials via capillary waves. With optical techniques such as fluorescent labelling or atomic force microscopy one also should be able to observe whether the action potential rides on a capillary carrier wave with a higher frequency, depending of the mode index n>1 of the utilized whispering gallery mode and the axon diameter via $n\lambda_T \simeq \pi D$. Note that for longitudinal wavelength similar to the tube diameter on expects a carrier frequency for capillary waves of order 100 kHz. From the low attenuation of ultrasound bulk waves in water in this frequency range one can also expect low damping for higher frequency capillary waves.

6. The neuron as a field-effect-tunable dipole transistor

The microscopic model discussed in the previous section can be understood as hydrodynamic [NEMS](), or more specifically a hydrodynamic dipolar Field-Effect-Transistor ([FET]()) in which the transport of signals from the soma acting as source to the synapsis acting as drain occurs through water waves with a field-effect-induced and -tunable polarization surrounding the perimeter adjacent to the lipid cell membrane. Here the lipid bilayer with the adjacent nanoscale layer of polarized water acts as a [dipolar]() [capacitor]() in which the density of water molecules oriented perpendicular to the lipid bilayer is controlled by the electric field in the high field region. At a finite voltage bias across the cell membrane the polarization density of the water molecules rises with increasing bias but will tend to saturate if all molecules in the high field layer will approach a completely polarized ferroelectric state. Hence the field-effect-induced polarization will change nonlinearly with voltage bias and may exhibit a threshold behavior when the originally unpolarized dielectric or antiferroelectric water makes a phase transition to a strongly polarized dielectric or ferroelectric state in the high field skin. Whereas the incompressible nature of liquid water keeps the density of the water molecules essentially constant the polarization of the electric-field-exposed region around the cell membrane behaves as a highly voltage-tunable liquid of partially oriented dipoles. Hence the mechanical radial force of polarized water depends on the induced polarization density and will also increase non-linearly with increasing bias voltage across the cell membrane. This electromechanical coupling is intrinsically caused by the polarization of water molecules. Since the total dipolar capacitance depends on the surface area of the lipid bilayer the resultant total polarization in the soma will be relatively large in comparison to the polarization of an axon section of similar length. Possibly this is why Nature has chosen to increase the surface area of the soma by dendritic branches as shown in Fig. 1, 6, and 10. Therefore any volume change in the soma will induce a change in the polarization skin of the axon and thus launch a polarization wave to propagate through the axon. Such a polarization wave packet may be caused not only by injection of ions into the soma that changes the local voltage bias but also by chemically or mechanically induced volume changes in the soma at a fixed finite voltage bias, the so called voltage clamp condition. Insofar propagation of action potentials can be caused by injection of ions into the soma that changes the local bias as described by the Hodgkin-Huxley model (4) or volume changes of the soma as described in the mechanical or electromechanical models (14 - 19, 29). For both cases the propagation of action potentials through the axon is likely to be dominated by polarization waves induced in the nanoscale layer of radially oriented water molecules. The propagation of such polarization wave packets in water thus is conceptually similar to the propagation of electromagnetic wave packets in field-effect transistors that are the working horse of todays digital electronics. This explains why both the phenomenological Hodgkin-Huxley model based on ionic currents and electric cable theory or the various electromechanical models discussed in detail in the review by Drukarch et al. (5) are able to describe the propagation of action potentials with suitably chosen fitting parameters. However, both lack to properly describe the underlying microscopic mechanisms. In this sense I find the here introduced field-effect-controlled hydrodynamic model based on the polarization of water dipoles that also is likely to support a coherent soliton-like propagation of action potentials more convincing. More studies building on previous experimental investigations exploring the electromechanical propagation of action potentials are needed to further support the here proposed microscopic mechanism of capillary wave transport of voltage-tunable polarization wave packets in liquid water.

7. Some further comments and speculative remarks.

When I first saw the picture of a myelinated axon on Wikipedia pictured above in Fig.6 I was wondering why Nature would choose such a chain-like construction. Thinking in terms of wave propagation I immediately thought of the cells between the nodes of Ranvier to look like cavities. Similar associations are triggered when one looks at their periodic arrangement in a chain reminiscent of one-dimensional phononic or photonic crystals. After all Nature has employed such cavity techniques for optical radiation to create beautifully colored butterfly wings by optical interference long before photonic crystals were created by physicist a few decades ago. This would imply that Nature has opened the toolbox of cavity opto- and nanomechanics for signal control by coherently coupling mechanical or electromechanical waves. In nanooptomechanics this topic that has caught strong recent interest (30,31) and has been utilized to engineer nanomechanical systems with fantastically high quality-factors in the range of $10^9$ at room temperature (13).

Mechanical waves with such large coherence enable access to the world of mechanically generated interference patterns similar to optical holograms. As [Chladni-Figures](#) such interference patterns generated by interference of mechanical waves in resonating plates are well-known. Similar interference patterns mechanically generated by water surface waves we can observe at any sandy beach. Analog to optical holograms one could envision non-local electromechanical storage of memories in the brain as mechanical holograms. Holograms store information in a spatially distributed interference pattern such that a local damage to a portion of the pattern does not severely reduce the information content stored. Possibly Nature has recognized this advantage and at least partially utilized electromechanical holography in our brain to store the huge amounts of information that we accumulate.

As the transverse whispering gallery modes can be considered as edge modes that can be decomposed into two counterpropagating chiral modes one is reminded of the lossless transport of edge modes of [topological insulators](#) caused by topological protection. In structural studies of myelinated axons via electron micrographs or optical [super-resolution microscopy](#) one has found spiral and periodic cylindrical structures with periodicities larger than 100 nm in the myelin sheath surrounding the axon ([32](#),[33](#)). This raises the question whether these serve to induce a screw-like sense of direction in the propagation of action potentials. Such a chiral component would cause some level of topological protection as it prevents backscattering. Such topological protection is not only essential for various versions of the [Quantum Hall Effect](#) but has also been recently demonstrated in the [Topomechanics group](#) at ETH Zurich to occur in macroscopic mechanic systems consisting of periodic arrays of coupled pendulums. Could it be that Nature has employed topological protection in nerve conduction long before it was discovered by physicist and honored with a Nobel prize in 2017?

With such speculations it seems possible that Nature has designed our nervous systems as a complex network of electromechanical signal propagation and storage that uses the outstanding features of wave mechanics in reduced dimensions that have formed the basis for quantum mechanics. I personally find it hard to believe that fragile quantum features such as [entanglement](#) have been used by Nature for storing and processing information ([34](#)). However, as I had the pleasure to learn with my collaborators for more than a decade by studying nanoelectromechanics, even the classical properties of wave mechanics can be outstanding and useful. In that context it also appears worthwhile to be attentive to what one presently learns in other collectively coupled quantum systems such as ultracold atoms and molecules in the growing field of [quantum hydrodynamics](#) as represented by recent studies of a quantum Newton's cradle ([35](#)).

In summary, I consider it essential to increase our study of nervous systems by including the possible role of guided electromechanical waves as coherent soliton-like wave packets and their potential for processing and storing information via interference phenomena. I hope that my remarks encourage younger colleagues to pursue experiments on cellular systems to study whether the transport of action potentials by the dipolar skin of capillary wave packets, propagating as soliton-like entities through axons with velocities that are easily predicted, pose a viable alternative to existing models of nervous signal transport and processing. Hopefully, such studies also increase the understanding of how our brain operates and clarify whether electromechanical wave interference could be a method chosen by Nature to store and process information.

Acknowledgements:

In the course of trying to find out whether an electromechanical model of propagation of action potentials is a viable alternative to existing models I profited enormously from discussions with colleagues and friends, by learning from their responses, constructive criticism, their questions that I was unable to answer promptly, and their guidance to existing literature that I was not yet aware of. Without their input I would not have been able to develop a hopefully somewhat coherent picture. I wish to thank all of them and in particular Matthias Schneider for making me aware of the controversial discussions and literature about how nerves are thought to operate, Achim Wixforth, Christoph Westerhausen, and Lukas Schnitzler for enthusiastically supporting my attemps to develop a reasonably consistent coupled electro-mechanical wave-based model and pointing me to literature that I had not be aware of, Thomas Heimburg for commenting on my efforts from his point of view, Jan Behrends, Niels Fertig, Peter Fratzl, Erwin Frey, Hermann Gaub, Alex Högele, Bert Lorenz, Florian Marquardt, Roland Netz, Michel Orrit, and Wilhelm Zwerger for stimulating discussions and hints, Chris Hohmann for supporting me in transforming my model into understandable graphics, and last not least my wife Sabine for her patience and understanding, when I started experiments with more or less confined water waves in our basement and was periodically absent from the real world with my thoughts about capillary wave transport and my Pilates-friends, who often had to listen to and question my thoughts in our weekly pub visit after our exhaustive physical exercise.

Financial support by the excellence cluster "Nanosystems Initiative Munich" (NIM) for supporting me as NIM senior member to travel to meetings and discussions with colleagues is gratefully acknowledged.

References:

(Note that the article title links to the journal page with abstract, whereas the reference number links to a full open access version of the article when available)